\documentclass[a4paper]{jpconf}
\usepackage{graphicx}
\begin{document}
\title{Searches for neutrinoless resonant double electron captures at LNGS}

\author{
P.~Belli$^{1}$,
R.~Bernabei$^{1,2}$,
R.S.~Boiko$^{3}$,
V.B.~Brudanin$^{4}$,
F.~Cappella$^{5,6}$,
V.~Caracciolo$^{7,8}$,
R.~Cerulli$^{7}$,
D.M.~Chernyak$^{3}$,
F.A.~Danevich$^{3}$,
S.~d'Angelo$^{1}$,
A.~Di~Marco$^{1}$,
M.L.~Di~Vacri$^{7}$,
E.N.~Galashov$^{9}$,
A.~Incicchitti$^{5,6}$,
V.V.~Kobychev$^{3}$,
G.P.~Kovtun$^{10}$,
N.G.~Kovtun$^{10}$,
V.M.~Mokina$^{3}$,
M.~Laubenstein$^{7}$,
S.S.~Nagorny$^{3}$,
S.~Nisi$^{7}$,
D.V.~Poda$^{3,7}$,
R.B.~Podviyanuk$^{3}$,
O.G.~Polischuk$^{3}$,
D.~Prosperi$^{5,6,}$\footnote[13]{Deceased.},
A.P.~Shcherban$^{10}$,
V.N.~Shlegel$^{9}$,
D.A.~Solopikhin$^{10}$,
Yu.G.~Stenin$^{9,13}$,
J.~Suhonen$^{11}$,
A.V.~Tolmachev$^{12}$,
V.I.~Tretyak$^{3}$,
Ya.V.~Vasiliev$^{9}$,
R.P.~Yavetskiy$^{12}$
}

\address{$^{1}$ INFN, Sezione di Roma ``Tor Vergata'', I-00133 Rome, Italy}
\address{$^{2}$ Dipartimento di Fisica, Universit\`a di Roma ``Tor Vergata'', I-00133 Rome, Italy}
\address{$^{3}$ Institute for Nuclear Research, MSP 03680 Kyiv, Ukraine}
\address{$^{4}$ Joint Institute for Nuclear Research, 141980 Dubna, Russia}
\address{$^{5}$ INFN, Sezione di Roma ``La Sapienza'', I-00185 Rome, Italy}
\address{$^{6}$ Dipartimento di Fisica, Universit\`a di Roma ``La Sapienza'', I-00185 Rome, Italy}
\address{$^{7}$ INFN, Laboratori Nazionali del Gran Sasso, I-67100 Assergi (Aq), Italy}
\address{$^{8}$ Dipartimento di Fisica, Universit\`a dell'Aquila, I-67100 L'Aquila, Italy}
\address{$^{9}$ Nikolaev Institute of Inorganic Chemistry, 630090 Novosibirsk, Russia}
\address{$^{10}$ Nat. Sci. Center ``Kharkiv Institute of Physics and Technology'', 61108 Kharkiv, Ukraine}
\address{$^{11}$ Department of Physics, University of Jyv$\ddot{a}$skyl$\ddot{a}$, P.O. Box 35 (YFL), FI-40014, Finland}
\address{$^{12}$ Institute for Single Crystals, 61001 Kharkiv, Ukraine}

\ead{rita.bernabei@roma2.infn.it}

\begin{abstract}
Several experiments were performed during last years at underground (3600 m w.e.)
Laboratori Nazionali del Gran Sasso (LNGS) of the INFN (Italy) to search for resonant
$2\varepsilon0\nu$ captures in $^{96}$Ru, $^{106}$Cd, $^{136}$Ce, $^{156}$Dy, $^{158}$Dy,
$^{180}$W, $^{184}$Os, $^{190}$Pt with the help of HP Ge semiconductor detectors,
and ZnWO$_4$ and $^{106}$CdWO$_4$
crystal scintillators. No evidence for r-$2\varepsilon0\nu$ decays was found,
and only $T_{1/2}$ limits were established in the range of $10^{14}-10^{21}$ yr.
\end{abstract}

\section{Introduction}

Resonant nuclear reactions occur sometimes in nature. Probably the most
prominent example of such a phenomenon is a triple $\alpha$ reaction.
There are no stable nuclei with $A=5$ and 8.
In these circumstances, the triple $\alpha$ reaction plays a critical role for
nucleosynthesis of heavier elements, when at first two $\alpha$ particles
create $^8$Be nucleus (with very short $T_{1/2} \simeq 10^{-16}$ s) and
one more $\alpha$ particle joins to create stable $^{12}$C.
Cross-section of  the $^8$Be + $\alpha$ reaction is not big enough to explain
abundances quantitatively. In 1953, F. Hoyle supposed \cite{Hoy53} that $^{12}$C
has an excited level at 7.68 MeV, and this should result in resonant enhancement
of the cross-section by orders of magnitude. This level was searched for and immediately observed
by experimentalists at $7.68\pm0.03$ MeV \cite{Dun53}.
It seems, that our world and all of us exist because of resonant enhancement of this nuclear reaction.

Resonant enhancement is expected also for neutrinoless double electron capture ($2\varepsilon0\nu$)
$e_1 + e_2 + (A,Z) \to (A,Z-2)$
in case of mass degeneracy of an initial and final (excited) nuclei:
\begin{center}
$Q = Q_{2\beta} - E_{b1} - E_{b2} = E_{exc}$,
\end{center}
where $Q_{2\beta} = \Delta M_a$ is the atomic mass difference between an initial and final nuclei.

The possibility of such resonant process (r-$2\varepsilon0\nu$) was discussed long ago \cite{th1},
where an enhancement of the rate by a few orders of magnitude was predicted
for perfect energy coincidence between the released energy and the energy of an excited state
($\simeq10$ eV). In this case the expected half lives could be as low as $\simeq10^{24}$ yr, and the corresponding
experiments even could compete with searches for neutrinoless $2\beta^-$ decay in sensitivity to the
neutrino mass. Theoretical aspects are discussed in recent papers \cite{th2}.

The first experimental limit on the r-$2\varepsilon0\nu$ process, to our knowledge, was
obtained for $^{106}$Cd $2K$ capture to $^{106}$Pd level with $E_{exc}=2741$ keV:
$T_{1/2}>3.0\times10^{19}$ yr at 90\% C.L. \cite{Bel99}. It was achieved in underground
measurements at LNGS of 154 g sample of Cd enriched in $^{106}$Cd to 68\% with two low background NaI(Tl)
scintillators working in coincidence. This enriched $^{106}$Cd isotope was used to grow
$^{106}$CdWO$_4$ crystal scintillator \cite{Bel10} for our further investigations of $2\beta$ processes in Cd and W.

Below, summary of our searches for r-$2\varepsilon0\nu$ captures in different nuclides
is given; the experiments were performed mainly during last two years.

\section{Experiments}

Different approaches were used in the experiments:
sharp peaks that correspond to $\gamma$ quanta emitted in deexcitation of $(A,Z-2)^*$ nuclei
were searched for r-$2\varepsilon0\nu$ processes in $^{96}$Ru, $^{136}$Ce, $^{156}$Dy, $^{158}$Dy,
$^{184}$Os, $^{190}$Pt with the help of HP Ge detectors,
while ZnWO$_4$ and $^{106}$CdWO$_4$ crystal scintillators
were applied in investigations of $^{180}$W and $^{106}$Cd, respectively.
Response functions and efficiencies were calculated with the GEANT4 simulation tool \cite{GEANT4};
initial kinematics of particles emitted in decays was generated with the DECAY0 event generator \cite{DECAY0}.

Natural isotopic abundance of $^{96}$Ru is $\delta=5.54\%$. Ru sample with mass
of 473 g was measured with HP Ge detector of 468 cm$^3$ during 158 h at the first step; then
data were collected also by a set-up with four HP Ge detectors ($\simeq 225$ cm$^3$ each) during 1176 h \cite{Bel09a}.
$T_{1/2}$ limits were established at the first time for r-$2\varepsilon0\nu$ in $^{96}$Ru at the level of $10^{19}$ yr
(see Table 1).
Quite strong pollution of the Ru sample by $^{40}$K at 3.4 Bq/kg was found \cite{Bel09a}. Additional purification
allowed to suppress $^{40}$K by one order of magnitude, and new data are under collection with the purified Ru sample
with mass of  $\simeq0.7$ kg.

For investigations of $^{106}$Cd, scintillating crystal $^{106}$CdWO$_4$ was developed \cite{Bel10}
with mass of 215 g and 66.4\% enrichment in $^{106}$Cd (while natural $\delta$ is 1.25\%).
The energy resolution of the detector is 10\% at 662 keV.
After 6590 h of data taking, new improved half life limits
on the r-$2\varepsilon0\nu$ processes were established at the level of up to $10^{21}$ yr \cite{Bel11a},
competitive or better than those set in the TGV experiment \cite{Ruk11}.

A sample of CeCl$_3$ crystal (6.9 g) was used to search for r-$2\varepsilon0\nu$ in $^{136}$Ce with HP Ge
244 cm$^3$ during 1280 h \cite{Bel09b}. $T_{1/2}$ limits were obtained for this nuclide
at the first time, but the small mass and poor natural abundance (0.185\%)
allowed to obtain the $T_{1/2}$ values on the level of only $10^{15}$ yr.

First searches for r-$2\varepsilon0\nu$ decays of $^{156}$Dy and $^{158}$Dy were performed with Dy$_2$O$_3$
sample with mass of 322 g (99.98\% purity grade) and HP Ge detector 244 cm$^3$ during 2512 h.
Once more, $T_{1/2}$ limits were not very big (around $10^{16}$ yr) because of low natural abundance of
$^{156}$Dy and $^{158}$Dy isotopes (0.056\% and 0.095\%, respectively) \cite{Bel11b}.

Several ZnWO$_4$ crystal scintillators with mass up to 699 g were used to search for r-$2\varepsilon0\nu$ decay of
$^{180}$W to the ground state of $^{180}$Hf. After near 19,000 h of measurements, the $T_{1/2}$ limit
was set as $1.3\times10^{18}$ yr \cite{Bel11c}.

For investigations of $^{184}$Os, sample of natural Os with mass of 172.5 g and purity grade of $>99.999\%$
(purified in the Kharkiv Institute of Physics and Technology, Ukraine; probably this is the most pure Os
in the world) is under measurements now with HP Ge 468 cm$^3$. The results are expected soon.

Natural platinum sample with mass of 42.5 g was measured with HP Ge detector 468 cm$^3$ during 1815 h.
The $^{190}$Pt isotope has quite low natural abundance of 0.014\% that leads to not very high
$T_{1/2}$ limit for r-$2\varepsilon0\nu$ decay of $2.9\times10^{16}$ yr \cite{Bel11e}.
As by-product of these measurements, also $\alpha$ decay of
$^{190}$Pt to the first excited level ($E_{exc}=137.2$ keV) of $^{186}$Os was observed at the first time;
corresponding half life is $T_{1/2} = 2.6^{+0.4}_{-0.3}(stat.)\pm0.6(syst.)\times10^{14}$ yr \cite{Bel11f}.

All the obtained results are summarized in Table 1 where also the expected energy releases $Q$ are compared
with the energies of excited levels of the final nuclei. 
It should be noted that the $Q$ values used in our papers on Ru, Ce, Dy were based on
the atomic masses \cite{Aud03} usually known with accuracy of a few keV.
Very recently many new, much more accurate results on the atomic masses appeared, 
motivated also by searches for perfect candidate for r-$2\varepsilon0\nu$ process.
The new values could be significantly shifted from the old ones sometimes removing an isotope
from the list of perspective r-$2\varepsilon0\nu$ candidates.
In particular, the old $Q_{2\beta}$ values \cite{Aud03} versus the new ones are the following:
for $^{96}$Ru  $2718\pm8$  vs. $2714.51\pm0.13$ keV \cite{Eli11a},
for $^{106}$Cd $2770\pm7$  vs. $2775.39\pm0.10$ keV \cite{Gon11},
for $^{136}$Ce $2419\pm13$ vs. $2378.53\pm0.27$ keV \cite{Kol11},
for $^{156}$Dy $2012\pm6$  vs. $2005.95\pm0.10$ keV \cite{Eli11b}.
However, one should remember also about possibility that some new excited
levels in daughter nuclei could be discovered with energies and $J^\pi$ values 
well fitted for the new atomic masses.

\footnotesize
\begin{table}
\caption{Summary of results on r-$2\varepsilon0\nu$ captures obtained in our experiments.
$T_{1/2}$ limits (yr) correspond to $90\%$ C.L. The $Q$ values were calculated with
the new atomic masses for $^{96}$Ru \cite{Eli11a}, $^{106}$Cd \cite{Gon11}, 
$^{136}$Ce \cite{Kol11}, $^{156}$Dy \cite{Eli11b}; for other nuclei masses of \cite{Aud03}
were used.}
\begin{center}
\begin{tabular}{lllllll}
\hline
Transition              &        & $Q$ (keV)   & $E_{exc}$ (keV), $J^\pi$ & Lim $T_{1/2}$         & Method           & Year               \\
\hline
$^{96}$Ru$\to^{96}$Mo   & $KL_1$ & 2691.6(1)   & 2700.2(1), $2^+$         & $2.2\times10^{19}$    & HP Ge            & 2009 \cite{Bel09a} \\
~                       & $2L_1$ & 2708.8(1)   & 2712.7(1)                & $5.1\times10^{19}$    & ~                & ~                  \\
$^{106}$Cd$\to^{106}$Pd & $2K$   & 2726.7(1)   & 2717.6(2)                & $4.3\times10^{20}$    & $^{106}$CdWO$_4$ & 2011 \cite{Bel11a} \\
~                       & $KL_1$ & 2747.4(1)   & 2741.0(5), $4^+$         & $9.5\times10^{20}$    & ~                & ~                  \\
~                       & $KL_3$ & 2747.9(1)   & 2748.2(4), $2,3^-$       & $4.3\times10^{20}$    & ~                & ~                  \\
$^{136}$Ce$\to^{136}$Ba & $2L_1$ & 2366.6(3)   & 2392.1(6), $(1^+,2^+)$   & $2.4\times10^{15}$    & HP Ge            & 2009 \cite{Bel09b} \\
~                       & $2L_1$ & 2366.6(3)   & 2399.9(1), $(1^+,2^+)$   & $4.1\times10^{15}$    & ~                & ~                  \\
$^{156}$Dy$\to^{156}$Gd & $2K$   & 1905.5(1)   & 1914.8(1), $2^+$         & $1.1\times10^{16}$    & HP Ge            & 2011 \cite{Bel11b} \\
~                       & $KL_1$ & 1947.3(1)   & 1946.4(1), $1^-$         & $9.6\times10^{15}$    & ~                & ~                  \\
~                       & $KL_1$ & 1947.3(1)   & 1952.4(1), $0^-$         & $2.6\times10^{16}$    & ~                & ~                  \\
~                       & $2L_1$ & 1989.2(1)   & 1988.5(2), $0^+$         & $1.9\times10^{16}$    & ~                & ~                  \\
~                       & $2L_3$ & 1991.5(1)   & 2003.7(1), $2^+$         & $3.0\times10^{14}$    & ~                & ~                  \\
$^{158}$Dy$\to^{158}$Gd & $2L_1$ & 268(3)      & 262,  $4^+$              & $3.2\times10^{16}$    & HP Ge            & 2011 \cite{Bel11b} \\
$^{180}$W$\to^{180}$Hf  & $2K$   & 13(4)       & g.s., $0^+$              & $1.3\times10^{18}$    & ZnWO$_4$         & 2011 \cite{Bel11c} \\
$^{184}$Os$\to^{184}$W  & $2K$   & 1312(1)     & 1322, $(0)^+$            & --                    & HP Ge            & 2011 \cite{Bel11d} \\
~                       & $KL_1$ & 1370(1)     & 1360, $(4^+)$            & --                    & ~                & ~                  \\
~                       & $2L_1$ & 1427(1)     & 1425, $(3)^+$            & --                    & ~                & ~                  \\
~                       & $2L_1$ & 1427(1)     & 1431, $2^+$              & --                    & ~                & ~                  \\
$^{190}$Pt$\to^{190}$Os & $2M$   & 1378(6)     & 1382, $(0,1,2)^+$        & $2.9\times10^{16}$    & HP Ge            & 2011 \cite{Bel11e} \\
\hline
\end{tabular}
\end{center}
\end{table}
\normalsize

\section{Conclusions}

Resonant $2\varepsilon0\nu$ captures in $^{96}$Ru, $^{106}$Cd, $^{136}$Ce, $^{156,158}$Dy, $^{180}$W, $^{190}$Pt
were searched for with HP Ge spectrometry and with scintillating crystals ZnWO$_4$ and $^{106}$CdWO$_4$.
Only $T_{1/2}$ limits were established from $3.0\times10^{14}$ to $9.6\times10^{20}$ yr.
These values are mostly the best today for these nuclides,
sometimes better than the previous ones by few orders of magnitude,
and sometimes they were obtained at the first time. However, they are still orders of magnitude worse than
those predicted by theory \cite{th1,th2}. An ``excellent'' candidate for r-$2\varepsilon0\nu$ is still not found.

In searches for r-$2\varepsilon0\nu$ captures, exact knowledge of $Q_{2\beta}$ values and $J^\pi$ properties
of excited levels is needed. Significant progress was reached recently for measurements of $Q_{2\beta}$ for
many $2\beta$ nuclides but we still need more accurate information for other isotopes.

Interesting by-products sometimes happen, like the first observation of $\alpha$ decay
$^{190}$Pt $\to$ $^{186}$Os$^*$ ($E_{exc}=137.2$ keV) with
$T_{1/2} = 2.6\times10^{14}$ yr \cite{Bel11f}. Experiments to search for resonant 
$2\varepsilon0\nu$ captures in $^{96}$Ru, $^{106}$Cd and $^{184}$Os are in progress at LNGS.

\ack
The work of the INR Kyiv group was supported in part by the Project
``Kosmomikrofizyka-2'' (Astroparticle Physics) of the National Academy
of Sciences of Ukraine.

\end{document}